\documentclass[14pt,preprint,groupedaddress]{revtex4}

\usepackage{graphicx}
\begin{document}


\title{Flashing Motor at High Transition Rate}


\author{Baoquan  Ai$^{a}$}\author{Liqiu Wang$^{a}$}\author{Lianggang
Liu$^{b}$}

\affiliation{$^{a}$ Department of Mechanical Engineering, The University of Hong Kong, Pokfulam Road, Hong Kong\\
 $^{b}$ Department of Physics, ZhongShan
University, GuangZhou, China}



\begin{abstract}
The movement of a Brownian particle in a fluctuating two-state
periodic potential is investigated. At high transition rate, we
use a perturbation method to obtain the analytical solution of the
model. It is found that the net current is a peaked function of
thermal noise, barrier height and the fluctuation ratio between
the two states. The thermal noise may facilitate the directed
motion at a finite intensity. The asymmetry parameter of the
potential is sensitive to the direction of the net current.
\end{abstract}

\pacs{05. 40. -a, 02. 50.Ey, 87. 10. +e}

\maketitle


\section{Introduction}
\indent

Much of the interest in non-equilibrium-induced transport
processes has been on the stochastically driven ratchets
\cite{1,2,3,4,5,6}. The noise-induced ratchet has recently
attracted considerable attentions. It is related to the symmetry
breaking and comes from the desire for an explaination of
directional transport in biological systems. Several models have
been, for example, proposed to describe the muscle contraction
\cite{7,8} and the asymmetric polymerization of action filaments
responsible for cell mobility \cite{9}.

 \indent
 The focus of research has been on the noise-induced unidirectional motion over the last decade. In these systems
directed-Brownian-motion of particles is generated by
nonequilibrium noise in the absence of any net macroscopic forces
and potential gradients. Ratchets have been proposed to model the
unidirectional motion due to the zero-mean nonequilibrium
fluctuation. Typical examples are rocking ratchets
\cite{10,11,12,13}, flashing ratchets \cite{14}, diffusion
ratchets \cite{15} and correlation ratchets \cite{16}. Ghosh et.
al.\cite{17,18} developed some analytical solutions of the current
variation with the noise in the ratchet. In all these studies the
potential is taken to be asymmetric in space. It has also been
shown that a unidirectional current can appear for spatially
symmetric potentials. For the case of spatially symmetric
potential, an external random force should be either temporally
asymmetric or spatially-dependent.

\indent The previous works are limited to case of single
potential. The present work extends the analysis to the case of
two potentials in flashing thermal ratchet: one constant potential
and one periodic in space and constant in time. No external
driving forces are required to induce unidirectional current in
these ratchets of two potentials. The emphasis is on the current
as the function of noise and other system parameters. This is
achieved by using a perturbation method to solve two coupled
Simoluchowsky equations.

\section{flashing motor}

\indent Consider the flashing motor (flashing ratchet) model
aiming for describing the spatially unidirectional motion along
x-direction in Fig. 1 of a Brownian particle due to the two
potentials (states): one constant and the other
spatially-periodic. This model was initially proposed in an
attempt of describing molecular motor in biological systems
\cite{19}.

 The rate of fluctuation between the two potential
states is governed by two rate constants $k_{1}$ and $k_{2}$,
respectively. Here the former is the rate from State 1 to State 2
and the latter is the rate from State 2 to State 1. While the
particle diffuses freely at State 2, it is localized near a local
minimum at State 1.  The particle motion satisfies the
dimensionless equation of motion
\begin{equation}\label{a}
    \frac{dx}{dt}=-\frac{V_{i}(x)}{dx}+f_{Bi}(t), i=1,2,
\end{equation}
where $f_{Bi}(t)$ is the Brownian random force, $x$ the position
of the particle, $t$ the time, $V_{i}(x)$ the potential, the
subscript $i$ stands for the state which can take value of 1 or 2.
The probability densities $P_{i}(x,t)(i=1,2)$ of state 1 and 2 are
govern by two coupled Simoluchowsky equations \cite{2}:
\begin{equation}\label{1}
\frac{\partial P_{1}(x,t)}{\partial t}=D\frac{\partial}{\partial
x}[\frac{\partial P_{1}(x,t)}{\partial
  x}-\frac{f(x)}{k_{B}T}P_{1}(x,t)]+k_{2}P_{2}(x,t)-k_{1}P_{1}(x,t)=-\frac{\partial J_{1}}{\partial x},
\end{equation}
\begin{equation}\label{2}
\frac{\partial P_{2}(x,t)}{\partial t}=D\frac{\partial^{2}
P_{2}(x,t)}{\partial
x^{2}}+k_{1}P_{1}(x,t)-k_{2}P_{2}(x,t)=-\frac{\partial
J_{2}}{\partial x},
\end{equation}
where $f(x)=-V^{'}_{1}(x)$, the prime stands for the derivative
with respect to $x$, $J_{1},J_{2}$ are probability densities of
current, $D$ the diffusivity, $k_{B}$ Boltzmann constant, $T$ the
absolute temperature. Here $x,t,k_{1},k_{2},D,k_{B}T$ are all
dimensionless. Also,
\begin{figure}[htbp]
  \begin{center}\includegraphics[width=10cm,height=8cm]{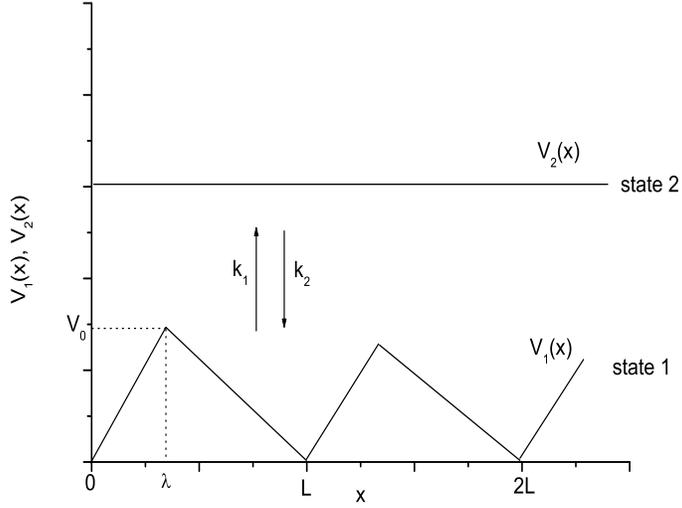}
  \caption{Two-state model with regular ratchets: $V_{1}(x)$ is spatially-periodic sawtooth
   of period $L$ and barrier height $V_{0}$. $\lambda$ is an asymmetry parameter ($V_{1}(x)$is symmetric when
   $\lambda=L/2$);
   $V_{2}(x)$ is a constant potential.}\label{1}
\end{center}
\end{figure}

\begin{equation}\label{}
 V_{1}(x)=\left\{
\begin{array}{ll}
    \frac{V_{0}}{\lambda}(x-mL),& \hbox{$mL<x\leq mL+\lambda$};\\
    \frac{V_{0}}{L-\lambda}[-x+(m+1)L],&\hbox{$mL+\lambda<x\leq(m+1)L$},\\
\end{array}
\right.
\end{equation}
where $m=0,1,2,...$. At steady state such that $\frac{\partial
P_{1}}{\partial t}=0$ and $\frac{\partial P_{2}}{\partial t}=0$,
Eqs (2) and (3) lead to the net current
\begin{equation}\label{3}
    J=-D\frac{\partial P(x)}{\partial x}+D_{1}f(x)P_{1}(x),
\end{equation}
\begin{equation}\label{4}
    D[P^{''}(x)-P_{1}^{''}(x)]+k[(1+\mu)P_{1}(x)-\mu P(x)]=0,
\end{equation}
where
\begin{equation}\label{}
J=J_{1}+J_{2}, P(x)=P_{1}(x)+P_{2}(x), D_{1}=\frac{D}{k_{B}T},
\end{equation}
and
\begin{equation}\label{}
k=k_{1}, \mu=k_{2}/k.
\end{equation}

\section{analytical solution}
When the fluctuation is at high rate such that $k>>1$, we can
expand $P(x)$, $P_{1}(x)$ and $J$ in power series of a small
parameters $k^{-1}$\cite{20},
\begin{equation}\label{5}
P(x)=\sum^{\infty}_{n=0}k^{-n}p_{n}(x),
P_{1}(x)=\sum^{\infty}_{n=0}k^{-n}p_{1n}(x),
J=\sum^{\infty}_{n=0}k^{-n}j_{n}.
\end{equation}
\indent The coefficients $p_{n}$, $p_{1n}$ and $j_{n}$ can be
obtained by substituting Eq.(9) into Eqs (5) and (6) and equating
coefficients of $k^{-n}$,
\begin{equation}\label{6}
    p_{0}^{'}(x)-\frac{\mu D_{1}}{(1+\mu)D}f(x)p_{0}(x)=-\frac{j_{0}}{D},
\end{equation}

\begin{equation}\label{7a}
p_{10}(x)=\frac{\mu}{1+\mu}p_{0}(x),
\end{equation}
\begin{equation}\label{7}
    -Dp_{n}^{'}(x)+\frac{\mu D_{1}}{1+\mu}f(x)p_{n}(x)=j_{n}+G_{n-1}(x),
    n=1,2,3,...,
\end{equation}
\begin{equation}\label{8}
G_{n}(x)=\frac{DD_{1}}{1+\mu}f(x)[p_{n}^{''}(x)-p_{1n}^{''}(x)],
n=0,1,2,....
\end{equation}
Under the periodicity conditions
\begin{equation}\label{9}
    p_{n}(x+L)=p_{n}(x), n=0,1,2,...
\end{equation}
and the normalization of the distribution $p(x)$ over the period
$L$,
\begin{equation}\label{10}
\int_{0}^{L}p_{n}(x)dx=\delta_{0n}, n=0,1,2,...,
\end{equation}
we can obtain all coefficients of $p_{n}$, $p_{1n}$ and $j_{n}$.
Since our attention is mainly on the current $J$, we only list
$j_{n}(x)$ here
\begin{eqnarray}\label{12}
\nonumber
j_{0}&=&0\\
j_{n}&=&-\frac{\int_{0}^{L}G_{n-1}(x)U^{-1}(x)dx}{\int_{0}^{L}U^{-1}(x)dx},
n=1,2,...
\end{eqnarray}
where
\begin{equation}\label{}
U(x)=\exp[-\frac{\mu D_{1}}{(1+\mu)D}V_{1}(x)].
\end{equation}

\indent In particular,
\begin{equation}\label{13}
j_{1}=-\frac{\mu^{2}D_{1}^{3}}{(1+\mu)^{4}D}\frac{\int_{0}^{L}f^{3}(x)dx}{\int_{0}^{L}U(x)dx\int_{0}^{L}U^{-1}(x)dx}.
\end{equation}
Therefore, to the first-order approximation,
\begin{equation}\label{14}
J\simeq
j_{0}+k^{-1}j_{1}=-\frac{\mu^{2}D_{1}^{3}}{k(1+\mu)^{4}D}\frac{\int_{0}^{L}f^{3}(x)dx}{\int_{0}^{L}U(x)dx\int_{0}^{L}U^{-1}(x)dx}.
\end{equation}
After substituting $f(x)$ and $U(x)$, we have
\begin{equation}\label{16}
J\simeq-\frac{\mu^{4}D^{2}V_{0}^{5}(2\lambda-L)}{(1+\mu)^{6}\beta^{5}kL\lambda^{2}(L-\lambda)^{2}(e^{\frac{\mu}{(1+\mu)\beta}V_{0}}+e^{-\frac{\mu}{(1+\mu)\beta}V_{0}}-2)},
\end{equation}
where $\beta=k_{B}T$. By letting $\frac{\partial J}{\partial
\gamma}=0$ with $\gamma=\frac{\beta}{V_{0}}$, we have
\begin{equation}\label{}
    (5\gamma-\frac{\mu}{1+\mu})\exp[\frac{\mu
    }{(1+\mu)\gamma}]+(5\gamma+\frac{\mu}{1+\mu})\exp[\frac{-\mu}{(1+\mu)\gamma}]-10\gamma=0,
\end{equation}
which leads to the optimum $\gamma$ for the maximum $J$ ($J_{max,
\gamma}$).

By letting $\frac{\partial J}{\partial \mu}=0$, similarly, we have
the optimum $\mu$ for the maximum $J$ ($J_{max,\mu}$),
\begin{equation}\label{}
    [-2\mu^{2}+(2-\frac{1}{\gamma})\mu+4)]\exp[\frac{\mu}{(1+\mu)\gamma}]+[-2\mu^{2}+(2+\frac{1}{\gamma})\mu+4)]\exp[\frac{-\mu}{(1+\mu)\gamma}]+4(\mu^{2}-\mu-2)=0.
\end{equation}

\section{Results and Discussion}
\indent Equation (20) indicates that the direction of the net
current is determined by the asymmetry parameter $\lambda$. When
$0<\lambda<1$,  the current is positive, the current is negative
when $1<\lambda<2$. There is no current at $\lambda=1$.
\begin{figure}[htbp]
  \begin{center}\includegraphics[width=11cm,height=8cm]{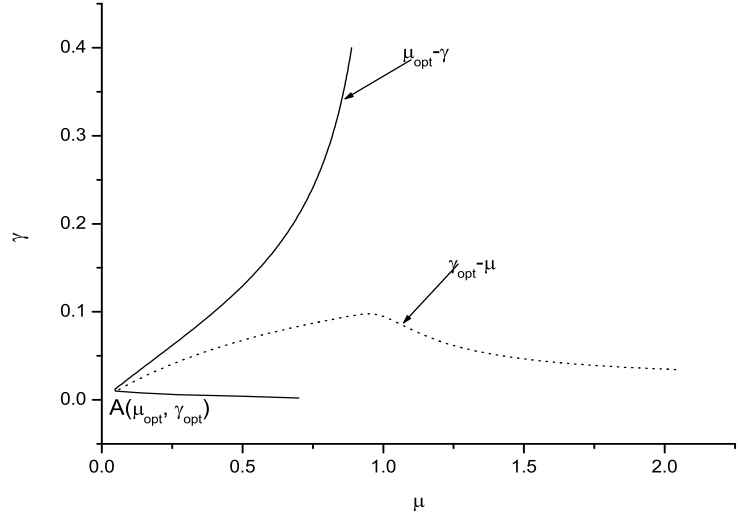}
  \caption{Plot of the optimum $\gamma$ vs the optimum $\mu$ for the maximum $J$ ($V_{0}=5.0$, $D=1.0$, $L=2.0$, $k=100.0$, $\lambda=0.5$).}\label{3}
\end{center}
\end{figure}

\begin{figure}[htbp]
  \begin{center}
\includegraphics[width=8cm,height=6cm]{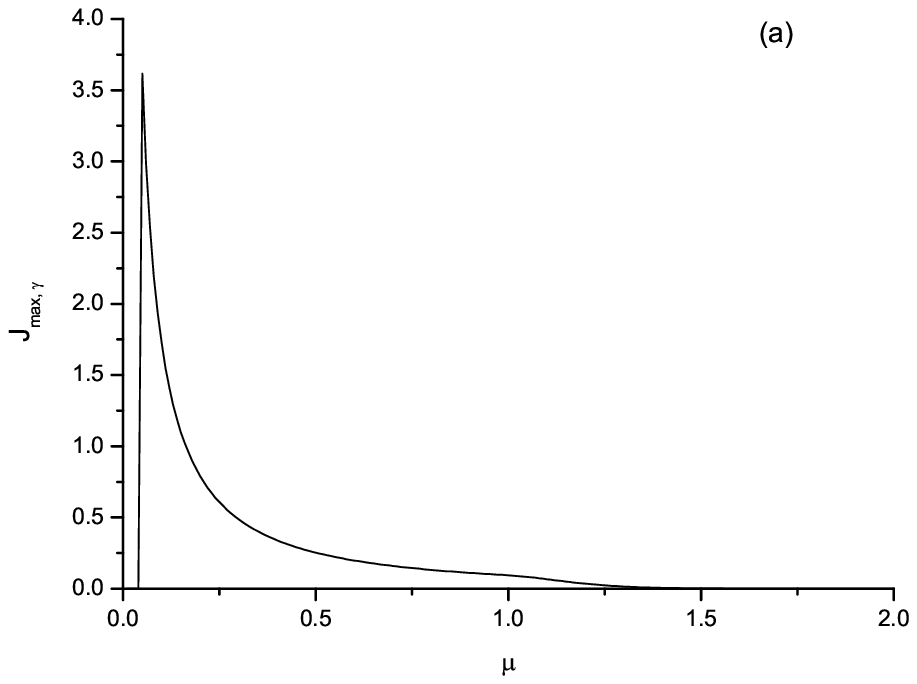}
\includegraphics[width=8cm,height=6cm]{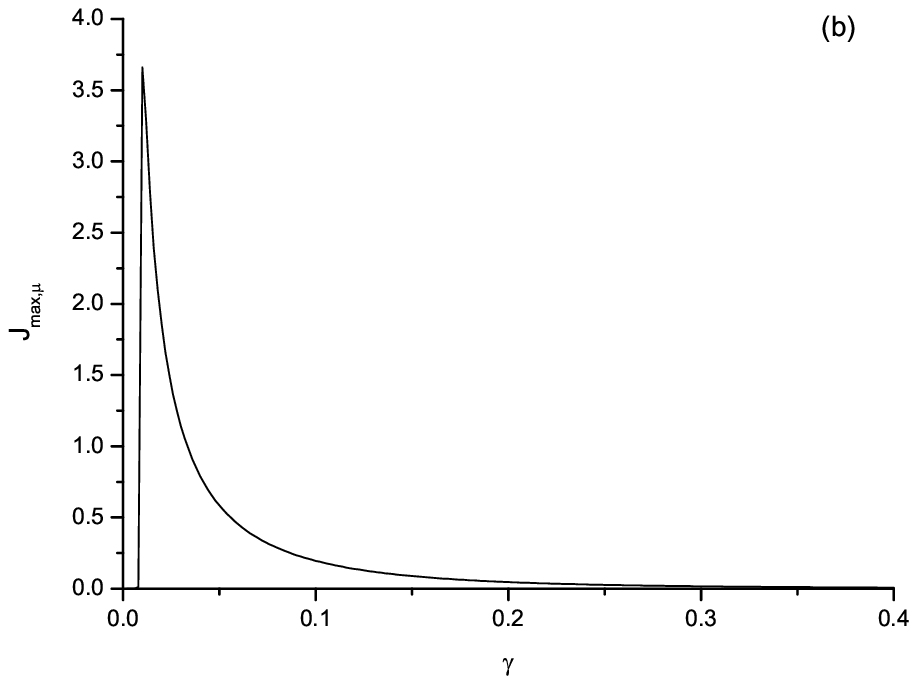}
  \caption{Plot of the maximum $J$ shown in Eqns. (21) and (22). (a) Plot of the maximum $J$ with the optimum $\gamma$
  vs $\mu$. (b) Plot of the maximum $J$ with the optimum $\mu$
  vs $\gamma$ ($V_{0}=5.0$, $D=1.0$, $L=2.0$, $k=100.0$, $\lambda=0.5$).}\label{3}
\end{center}
\end{figure}
\indent  Figure 2 shows the solution of the Eqns (21) and (22).
The dot line gives the optimum $\gamma$ for the maximum $J$ (Eq.
(21)) and the solid line gives the optimum $\mu$ for the maximum
$J$ (Eq. (22)). It is easy to find that the dot line meets the
solid line at the point $A$ ($\mu_{opt}$,$\gamma_{opt})$, which
indicates that one can obtain the maximum $J$ for both the optimum
$\gamma$ and the optimum $\mu$ at the same time. The corresponding
$J_{max, \gamma}$ vs $\mu$ and $J_{max,\mu}$ vs $\gamma$ are shown
in Fig. 3a and Fig. 3b, respectively. From Fig. 3a, we can find
$J_{max, \gamma}$ as the function of $\mu$ have a maximum value,
at which the $\mu$ and $\gamma$ are optimal, namely, the solid
line will meet the dot line as shown in Fig. 2. The similar
results can also be obtained in Fig. 3b.

\begin{figure}[htbp]
  \begin{center}\includegraphics[width=11cm,height=8cm]{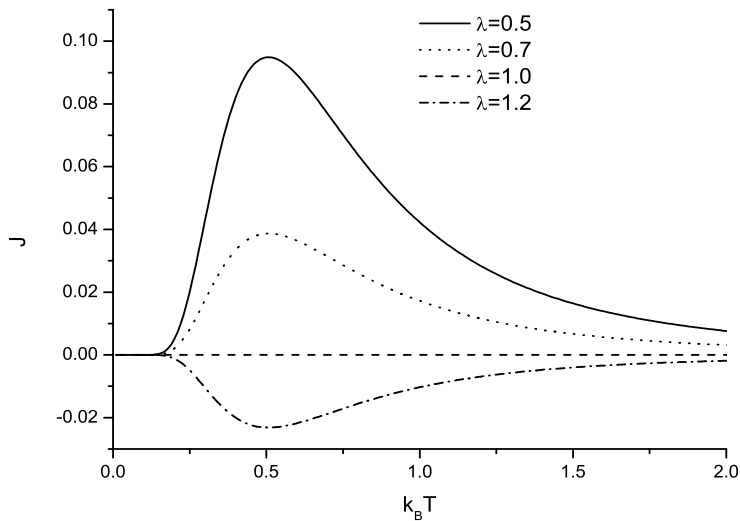}
  \caption{Dimensionless probability current $J$ vs thermal noise strength $k_{B}T$ for different values of asymmetric parameters
  ($V_{0}=5.0$, $D=1.0$, $L=2.0$, $k=100.0$, $\mu=1.0$).}
  \label{2}
\end{center}
\end{figure}

\indent Figure 4 shows the variation of the net current $J$ with
the thermal noise intensity $k_{B}T$. The curve is observed to be
bell-shaped, a feature of resonance. The current reversal appears
at $\lambda=1$ at which the potential $V_{1}(x)$ is symmetry. When
$k_{B}T\rightarrow 0$, $J$ tends to zero for all values of
$\lambda$. Therefore, there are no transitions out of the wells
when the thermal noise vanishes. When $k_{B}T\rightarrow \infty$
so that the thermal noise is very large, the ratchet effect also
disappear. The current $|J|$ has a maximum value for fixed value
$\lambda$ at certain value of $k_{B}T$. By Eq. (21), the optimized
value is $k_{B}T=0.5073$($\gamma=0.1015, V_{0}=5.0$). The maximum
$J_{max}=0.0949$ at $\lambda=0.5$. Therefore, certain thermal
noise can induce a large current $|J|$, while the thermal noise
blocks the unidirectional motion in general.

\begin{figure}[htbp]
  \begin{center}\includegraphics[width=11cm,height=8cm]{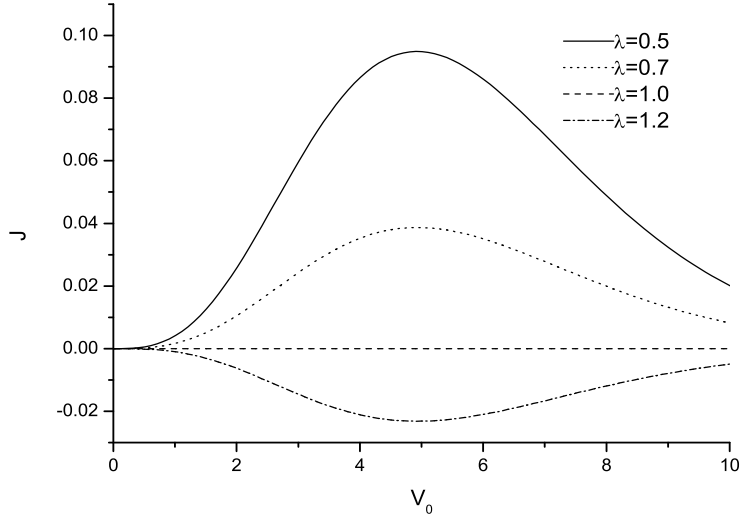}
  \caption{Dimensionless probability current $J$ vs barrier height $V_{0}$ for different values of $\lambda$ ($k_{B}T=0.5$, $D=1.0$, $L=2.0$, $k=100.0$, $\mu=1.0$).}\label{3}
\end{center}
\end{figure}

\indent Figure 5 shows the net current as the function of barrier
height $V_{0}$. We again observe the current reversals at
$\lambda=1.0$. When the barrier height $V_{0}$ is small. The
effect of the ratchet is also small; the thermal noise effect is
domaint so that the net current disappear. When the barrier height
$V_{0}$ is large, on the other hand, the particle can not pass the
barrier. It can only diffuse at State 1 so that
 the net current is also very small. Therefore, there is an
 optimized value of $V_{0}$ (4.9281) at which $J$ takes its maximum value($J_{max}=0.0949$), for example, $\lambda=0.5$.


\begin{figure}[htbp]
  \begin{center}\includegraphics[width=11cm,height=8cm]{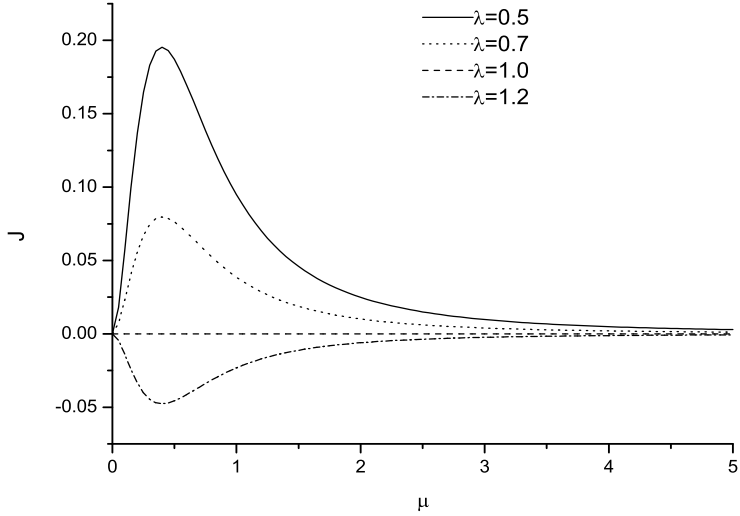}
  \caption{Dimensionless probability current $J$ vs the ratio of the two state transition rate $\mu$ for different values of
  $\lambda$ ($k_{B}T=0.5$, $V_{0}=5.0$, $D=1.0$, $L=2.0$,
$k=100.0$).}\label{3}
\end{center}
\end{figure}
\indent Figure 6 shows the current as the function of $\mu$. When
$\mu\rightarrow 0$, $k_{2}$ tends to zero. The attraction by State
2 is too small such that the particle is always staying at State
1. The ratchet reduces to one-state ratchet without external
force. Therefore, no current exits. When $\mu\gg 1$, the
attraction by State 1 becomes too small so that the particle can
only diffuse at State 2. The net current becomes to zero again.
Hence, there exits an optimized value of $\mu$ (0.3991) at which
$J$ takes its maximum value($J_{max}=0.1954$) as shown in Eq.
(22).

\section{Concluding Remarks}
\indent  Two coupled Simoluchowsky equations are solved by a
perturbation method to obtain the net current. The current is
peaked function of thermal noise $k_{B}T$, barrier height $V_{0}$
and the fluctuation ratio $\mu$ between the two states. It is
positive for $0<\lambda<1$ and negative for $1<\lambda<2$.
Therefore, the current reverses its direction at $\lambda=1.0$
(symmetric potential). When the thermal noise is small, the
particle can not pass the barrier such that the current $J$ tends
to zero. When the thermal noise is too large, the ratchet effect
disappear so that $J$ tends to zero, also. There is an optimized
value of thermal noise at which $J$ takes its maximum value. For
the case of $V_{0}\rightarrow 0$, the thermal noise is dormant and
the current disappears. When $V_{0}\rightarrow\infty$, on the
other hand, the particle can not pass the barrier. When
$\mu\rightarrow 0$, the attraction from State 2 is too small and
the particle is always at State 1. The ratchet reduces to
one-state ratchet without external force. Therefore, no currents
occur. When $\mu$ is very large, similarly, the attraction from
State 1 is too small, the particle can only stay at State 2, and
the current also tends to zero. There exits optimized values of
$k_{B}T$,$V_{0}$ and $\mu$ at which the current takes its maximum
value.

\indent Here, the thermal noise can facilitate the directed motion
of the Brownian particles. This differs from the one prevalent in
the literature that the thermal noise always destroyed the
directed motion. The noise-induced transport is associated with
the breaking of either reflection symmetry of spatially periodic
system or statistical symmetry of temporal nonequilibrium
fluctuations characterized by multitime correlation functions. The
symmetry-breaking driven transport can be used to explain the
directed motion of macromolecules in biological cell and to
construct well-controlled devices of high resolution for
separation of macro-particles and micro-particles \cite{21}.\\

\end{document}